\documentstyle[aps,prl,multicol,psfig,floats,fancyheadings]{revtex}
\begin{document}
\draft
\twocolumn[\hsize\textwidth\columnwidth\hsize\csname@twocolumnfalse\endcsname
\title{ Electron Localization in the Insulating State } \author{Raffaele
Resta} \address{Istituto Nazionale di Fisica della Materia (INFM), Strada
Costiera 11, I--34014 Trieste, Italy \\ Dipartimento di Fisica Teorica,
Universit\`a di Trieste, I--34014 Trieste, Italy} \author{Sandro Sorella}
\address{Istituto Nazionale di Fisica della Materia (INFM), Via Beirut 4,
I--34014 Trieste, Italy \\ Scuola Internazionale Superiore di Stud\^\i\
Avanzati (SISSA), Via Beirut 4, 34014 Trieste, Italy}

\date{November 2, 1998} 
\maketitle

\begin{abstract} The insulating state of matter is characterized by the
excitation spectrum, but also by qualitative features of the electronic
ground state. The insulating ground wavefunction in fact: (i) sustains
macroscopic polarization, and (ii) is {\it localized}. We give a sharp
definition of the latter concept, and we show how the two basic features stem
from essentially the same formalism. Our approach to localization is
exemplified by means of a two--band Hubbard model in one dimension. In the
noninteracting limit the wavefunction localization is measured by the spread
of the Wannier orbitals.  \end{abstract}

\bigskip\bigskip

]
\narrowtext

In a milestone paper appeared in 1964~\cite{Kohn64} W. Kohn investigated the
very basic features which discriminate between an insulator and a metal: he
gave evidence that {\it localization} of the electronic ground wavefunction
implies zero DC conductivity, and therefore characterizes the insulating
state. In this Letter we provide a definition of localization which is
deeply rooted into the modern theory of
polarization~\cite{modern,rap_a12,Ortiz94,rap100}, and rather different from
Kohn's one. Indeed, besides zero DC conductivity, the property which
obviously discriminates between insulators and metals is dielectric
polarization: whenever the bulk symmetry is low enough, an insulator
displays nontrivial static polarization. Here we show that the whole
information needed for describing {\it both} localization and polarization
is embedded into the same many--body expectation value: namely, the complex
number $z_N$ defined in Eq.~(\ref{general}) below. It was previously
shown~\cite{rap100} that macroscopic polarization is essentially the {\it
phase} of $z_N$: here we show that the {\it modulus} of $z_N$ yields a
definition of localization length which is sharper and more meaningful than
the available ones. In our formalism a vanishing $z_N$ implies a delocalized
wavefunction {\it and} an ill--defined polarization: this characterizes the
metallic state. Our definition is first demonstrated for a one--dimensional
crystalline system of independent electrons, in which case our localization
length coincides (for insulators) with the spread of the Wannier orbitals.
We then study a two--band Hubbard model undergoing a Mott-like transition:
both in the band regime (below the transition) and in the highly correlated
regime (above the transition) the wavefunction turns out to be localized,
while the localization length diverges at the transition point, thus
indicating a metallic ground state.  Our approach to localization in a
many--electron system sharply discriminates between a conducting and
nonconducting ground state, yet avoiding any reference to the excitation
spectrum.

Let us start with a single one--dimensional electron: the distinction between
localized (bound) and delocalized (scattering) states is a clearcut one when
the usual boundary conditions are adopted; much less so when periodic
Born--von--K\`arm\`an boundary conditions (BvK) are adopted, implying a ring
topology for the one--dimensional system.  Within the latter choice---which
is almost mandatory in condensed matter physics---all states appear in a
sense as ``delocalized'' since all wavefunctions $\psi(x)$ are periodic over
the BvK period: $\psi(x+L) = \psi(x)$.  
\begin{figure}  \centerline{\psfig{file=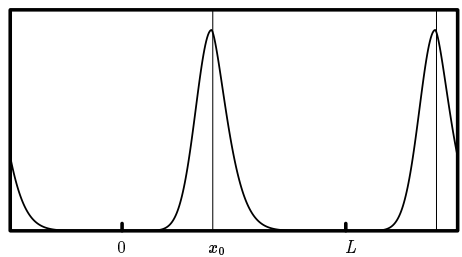,width=6cm}}
\caption[1]{The distribution $|\psi(x)|^2$ of a localized state
within periodic Born--von--K\`arm\`an boundary conditions}
\label{loc} \end{figure}
We are going to show that the key
parameter to study localization of an electronic state within BvK is the
dimensionless complex number $z$, defined as \begin{equation} z = \int_0^L
\!\! dx \;  {\rm e}^{i\frac{2\pi}{L} x} |\psi(x)|^2 , \end{equation} whose
modulus is no larger than 1.  In the case of extreme delocalization one has
$|\psi(x)|^2 = 1/L$ and $z = 0$, while in the case of extreme localization
\begin{equation} |\psi(x)|^2 = \sum_{m = -\infty}^\infty \delta(x - x_0 -mL)
, \end{equation} and we get $z = {\rm e}^{i\frac{2\pi}{L} x_0}$. In the most
general case, depicted in Fig.~\ref{loc}, the electron density $|\psi(x)|^2$
can always be written as a superposition of a function $n_{\rm loc}$,
normalized over $(-\infty, \infty)$, and of its periodic replicas:
\begin{equation} |\psi(x)|^2 = \sum_{m = -\infty}^\infty n_{\rm loc} (x - x_0
-mL) .  \label{replicas} \end{equation} Both $x_0$ and $n_{\rm loc} (x)$ have
a large arbitrariness: we restrict it a little bit by imposing that $x_0$ is
the center of the distribution, in the sense that $\int_{-\infty}^\infty dx
\, x n_{\rm loc} (x) =0$.

Using Eq.~(\ref{replicas}), $z$ can be expressed in terms of the Fourier
transform of $n_{\rm loc}$ as: \begin{equation} z = {\rm e}^{i\frac{2\pi}{L}
x_0} \tilde{n}_{\rm loc} (-\frac{2\pi}{L}) . \end{equation} The distinction
between a localized and a delocalized state becomes clear if one studies the
behavior of $z$ when the BvK periodicity $L$ is varied. For a localized
state, in fact, the shape of $n_{\rm loc} (x)$ is essentially
$L$--independent (exponentially with $L$ for large $L$), while the opposite
is true for a delocalized state.  If the electron is localized in a region of
space much smaller than $L$, its Fourier transform is smooth over reciprocal
distances of the order of $L^{-1}$ and can be expanded as: \begin{equation}
\tilde{n}_{\rm loc} (-\frac{2\pi}{L}) = 1 - \frac{1}{2}
\left(\frac{2\pi}{L}\right)^2 \int_{-\infty}^\infty dx \, x^2 n_{\rm loc} (x)
+  {\cal O}(L^{-3}) \label{expansion} . \end{equation} Therefore at the
increase of $L$, $|z|$ tends to 1 for a localized state, while it vanishes in
the delocalized case.

A very natural definition of the center of a localized periodic distribution
$|\psi(x)|^2$ is provided by the phase of $z$ through the formula:
\begin{equation} \langle x \rangle = \frac{L}{2\pi} \mbox{Im ln}\,  z ,
\end{equation} first proposed by Selloni {\it et al.} in
Ref.~\onlinecite{heuris} to track the adiabatic time evolution of a single
quantum particle in a disordered condensed system within BvK.  The
expectation value $\langle x \rangle$ is defined modulo $L$, as expected
since $|\psi(x)|^2$ is periodic: the previous equations imply indeed
$\langle x \rangle \simeq x_0$ mod $L$.  The modulus of $z$ can be used to
measure the localization length $\lambda$.  Using Eq.~(\ref{expansion}) we
get \begin{equation} \ln |z| \simeq - \frac{1}{2}
\left(\frac{2\pi}{L}\right)^2 \int_{-\infty}^\infty dx \, x^2 n_{\rm loc}
(x) ,\end{equation} and the spread of the electronic distribution can be
defined through: \begin{equation} \lambda^2 = \langle x^2 \rangle - \langle
x \rangle^2 = - \left(\frac{L}{2\pi}\right)^2 \ln |z|^2 , \label{spread}
\end{equation} which for large $L$ goes to a constant limit for a localized
state, and diverges for a delocalized one. Eq.~(\ref{spread}) provides an
alternative measure of localization with respect to the usual participation
ratio\cite{participation}.

So much about the one--electron problem. We are now going to consider a
finite density of electrons $n_0$: $N$ particles in a periodic box of size
$L$.  Eventually, the thermodynamic limit is taken: $N \rightarrow \infty$,
$L \rightarrow \infty$, $N/L = n_0$ constant. Even for a system of
independent electrons, our approach takes a simple and compact form if a
many--body formulation is adopted. In this case the ground state obeys BvK
in each electronic variable separately: \begin{equation} \Psi(x_1, \dots,
x_i, \dots, x_N) = \Psi(x_1, \dots, x_i\! + \! L, \dots, x_N) . 
\label{perio} \end{equation} Spin variables are not explicitated (here and
in the following formulas), while of course are taken care of in the
calculations. In analogy with the one--particle case, we define the
many--body multiplicative operator $\hat{X} = \sum_{i=1}^N x_i$, and the
complex number \begin{equation} z_N = \langle \Psi | {\rm
e}^{i\frac{2\pi}{L} \hat{X}} | \Psi \rangle , \label{general} \end{equation}
which will be used to discriminate between a localized many--body ground
eigenstate (where $|z_N| \rightarrow 1$ for large $N$) and a delocalized
one, where $z_N$ vanishes. Ergo, following Kohn's viewpoint~\cite{Kohn64},
the modulus of $z_N$ will be used here to discriminate between insulators
and metals. We start with the dimensionless quantity \begin{equation} D = -
\lim_{N \rightarrow \infty} N \ln |z_N|^2, \label{local} \end{equation}
which is finite in insulators and divergent in metals: we define the
localization length as $ \lambda = \sqrt{D}/(2 \pi n_0)$.  We emphasize that
our definition of localization---as well as the definition of polarization
given in Ref.~\cite{rap100}---deals on the same ground with a general
system, either ordered or disordered, either independent--electron or
correlated.

For a crystalline system of independent electrons the many-body wavefunction
$\Psi$ can be written as a Slater determinant of Bloch orbitals and $z_N$
factorizes. Using the same algebra as in Ref.~\onlinecite{rap100}, Eq.~(14)
onwards, one can easily prove that for a metal $z_N$ vanishes, while for an
insulator $D$ converges to the Brillouin--zone (BZ) integral:
\begin{eqnarray} D = 4 m_b \frac{2\pi}{a} \int_{\rm BZ} \!\! dk \; [
\sum_{m=1}^{m_b} \langle u'_{m,k} | u'_{m,k} \rangle \nonumber \\ -
\sum_{l,m=1}^{m_b} \langle u'_{m,k} | u_{l,k} \rangle \langle u_{l,k} |
u'_{m,k} \rangle ] .  \label{distance} \end{eqnarray} In Eq.~(\ref{distance})
we assume a linear system of lattice constant $a$ with $m_b$ occupied bands
and density $n_0\!=\!2 m_b/a$; $u_{m,k}$ is the periodic factor in the Bloch
orbital (chosen to be a differentiable function of $k$), and the prime
indicates the $k$-derivative.  The integral in Eq.~(\ref{distance}) is a
``geometric distance''~\cite{Pati91} and measures the spread $\lambda^2 =
\langle x^2 \rangle - \langle x \rangle^2$ of the optimally localized Wannier
orbitals~\cite{Kohn59,Marzari97}: our $\lambda^2$ coincides in fact with
$\Omega_{\rm I}/m_b$, where $\Omega_{\rm I}$ is the quantity defined by
Marzari and Vanderbilt~\cite{nota} (in one dimension).  

Next we study a one--dimensional model of a correlated polar crystal. We
focus on the centrosymmetric case where $z_N$ is real and its phase is
either 0 or $\pi$.  We choose a two--band Hubbard model at half filling,
whose Hamiltonian is: \begin{equation} \sum_{j \sigma} [ \; (-1)^j \Delta \,
c^\dagger_{j \sigma} c_{j \sigma} - t ( c^\dagger_{j \sigma} c_{j+1 \sigma}
+ \mbox{\rm H.c.} ) \; ] + U \sum_j n_{j\uparrow} n_{j\downarrow} \; ,
\label{hubbard} \end{equation} and depends on two parameters besides the
Hubbard $U$: the hopping $t$, and  the difference in site energies $2
\Delta$.  

\begin{figure}  \centerline{\psfig{file=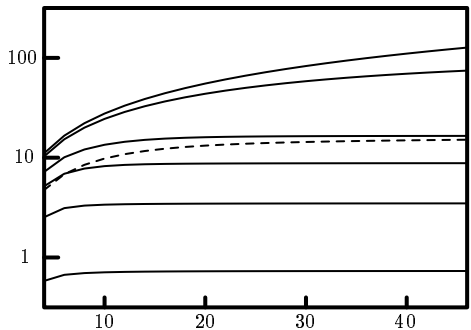,width=6cm}}
\caption[1]{Solid lines: $-N \ln |z_N|^2$ as a function of $N$
at several values of $\Delta/t$ for independent electrons ($U=0$). From
bottom to top, $\Delta/t$ assumes the values: 5, 2, 1, 0.5714, 0.1, 0.01,
respectively. The dashed line is obtained replacing the logarithm with its
leading expansion at $\Delta/t= 0.5714$.}
\label{conv} \end{figure}

In the special case $U\!=\!0$  we recover a system of independent electrons
and the model describes an insulator whenever $\Delta \! \neq \! 0$. As
discussed above, $D$ is finite, Eq.~(\ref{distance}), in the insulating case
and formally infinite (even at {\it finite} $N$) in the metallic one. We
show in Fig.~\ref{conv} the convergence of $D$ for several values of
$\Delta/t$: the localization length diverges upon approaching the metallic
state ($\Delta=0$), and a large system size $N$ is needed for evaluating $D$
if $\Delta/t$ is small. Approximating $D$ with its finite--$N$ value
is exactly equivalent to a discretization of the BZ integral in
Eq.~(\ref{distance}): if we further replace the logarithm in
Eq.~(\ref{local}) with its leading expansion, we recover the same
discretization proposed in Ref.~\onlinecite{Marzari97}. While one gets the
same limiting value, our logarithm form converges much faster: this is also
shown in Fig.~\ref{conv}, for a selected value of $\Delta/t$.  

The case of $U\!>\!0$ is much more interesting, since no Wannier functions
or single--particle orbitals can be defined. Notwithstanding, our $\lambda$
mantains its value of a meaningful measure of the localization of the
many--body wavefunction as a whole, even in the highly correlated regime. 
The model Hamiltonian of Eq.~(\ref{hubbard}) has been thoroughly studied by
several authors \cite{Egami,rap87,Ortiz95}: when $U$ is increased to large
values, the system undergoes an interesting transition, from a band
insulator to a Mott insulator. The real number $z_N$ changes sign at the
transition: this fact has an important physical meaning, since it indicates
a swapping of roles between anion and cation. In the low $U$ regime the
anion is the ion having the lowest on--site energy (odd $j$ for positive
$\Delta$), while the opposite is true in the highly correlated regime.  We
have shown in Ref.~\onlinecite{rap87} that the anion--cation swap manifests
itself in a discontinuous change of the {\it dynamical} (or Born) ionic
charge, while instead the static charge is {\it continuous} and carries no
information about the transition. The many--body wavefunction is explicitly
needed for detecting the transition, and the relevant information is indeed
embedded in $z_N$. We adopt in the following the value of $\Delta/t =
0.5714$, previously used in Ref.~\onlinecite{rap87,Ortiz95}: the Mott
transition occurs then at $U=2.27t$.

\begin{figure}  \centerline{\psfig{file=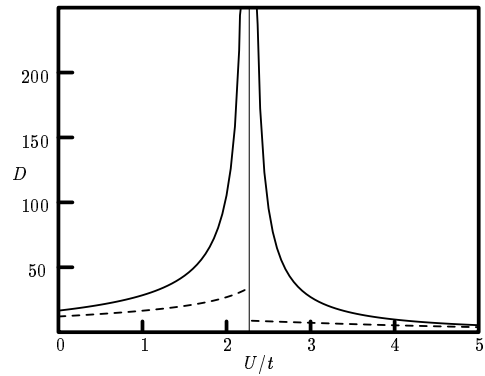,width=6cm}}
\caption[1]{Dimensionless localization parameter $D$,
Eq.~(\protect\ref{local}), as a function of $U/t$, where $\Delta/t= 0.5714$. 
Dashed line: calculations at $N=8$, which are not converged since the size is
too small. Solid line: values of $D$, where the thermodynamic limit is
achieved by means of the ansatz wavefunctions (see text); the divergence at
the Mott transition is perspicuous.}
\label{varu} \end{figure}

We perform exact diagonalizations for $N$=8 via the Lanczos algorithm, as
described in Ref.~\onlinecite{rap87}. The results are shown in
Fig.~\ref{varu}, dashed line, and would indicate an increase of the
localization length until the transition point, where a discontinuous drop
occurs; in the high--$U$ region the wavefunction is strongly localized. 
However, upon performing the calculations in this way the finite size effects
are clearly very relevant: this depends on the chosen value of $\Delta/t$. 
Even at $U$=0 (where we can afford exact diagonalizations at arbitrarily
large sizes) the value of $D$ calculated at $N$=8 differs from the fully
converged value by 27\% (see Fig.~\ref{conv}): in the correlated case the
situation is expected to worsen.  We have performed a few calculations at
different sizes, up to $N=12$: the convergence turns out to be slow and
oscillatory, with $N=4n$ and $N=4n+2$ following different trends.

We overcome this drawback upon building approximate wavefunctions for much
larger sizes. At a fixed size $N$ we perform {\it several} independent
calculations, using skew (quasiperiodic) boundary conditions with Bloch
vector $k$ over each electronic variable separately: \begin{equation}
\Phi_k(x_1, \dots, x_i\! + \! L, \dots, x_N) = {\rm e}^{ikL} \Phi_k(x_1,
\dots, x_i, \dots, x_N) \label{skew} .  \end{equation} We choose $M$ equally
spaced values of $k$ in the interval $[0, 2\pi/L)$: \begin{equation} k_s =
\frac{2 \pi}{M L} s, \quad s=0,1,\dots, M\! -\!1 . \label{points}
\end{equation} Each of the $\Phi_{k_{s}}$ is therefore BvK periodical over a
period $L' = ML$, and we build an ansatz wavefunction for $N' =MN$ electrons
as the antisymmetrized product of the $M$ different $N$--particle
wavefunctions $\Phi_{k_{s}}$. In the simple case of $N=1$ this construction
yields the Slater determinant of $M$ orbitals, and is therefore the exact
wavefunction for a system of $N'=M$ noninteracting electrons. Upon choosing
$N > 1$  one allows the $MN$ electrons to correlate, but only in clusters of
$N$ at a time: of course, our ansatz wavefunction has restricted variational
freedom.  At any given $M$, and for even $N$, the number $z_{N'}$ factorizes
as~\cite{nota2}: \begin{equation}z_{N'} = z_{MN}= \langle \Psi | {\rm
e}^{i\frac{2\pi}{ML} \hat{X}} | \Psi \rangle = \prod_{s=0}^{M-1} \langle
\Phi_{k_{s+1}} | {\rm e}^{i\frac{2\pi}{ML} \hat{X}} | \Phi_{k_{s}} \rangle .
\end{equation} For instance taking $N=4$ and $M=3$ the ansatz reproduces the
exact 12 sites result (at $U/t = 1$) within 3\%. We then approximate the
thermodynamic limit upon studying the large $M$ limit at fixed $N$:
\begin{equation} D = - \lim_{M \rightarrow \infty} NM \ln |z_{NM}|^2 . 
\end{equation}  The values of $D$ calculated for $M=100$ and $N=8$ are
plotted in Fig.~\ref{varu}, solid line: one clearly sees a divergence of the
localization length at the Mott transition, while the wavefunction becomes
localized again in the highly correlated regime.  

The many--dimensional generalization of the present formulation is not
straightforward: its presentation is outside the scope of the present
Letter. We mention here only a few main features. (i): $\lambda$ is
essentially a unidimensional quantity, in the sense that one fixes a
direction and defines a localization length in that given direction: say
$\lambda_{xx}$ if we choose $z_N$ identical in form to Eq.~(\ref{general}). 
For anisotropic crystals, different $\lambda$'s coexist: for instance in
graphite we expect $\lambda$ to be finite in the direction normal to the
basal plane, and divergent in the planar direction. (ii): For a
$d$-dimensional system of $N$ electrons in a cubic box of volume $L^d$ the
factor $N$ appearing in Eq.~(\ref{local}) must be replaced with $N^{2/d-1}$
in order to define the dimensionless $D_{xx}$, and the localization length
becomes: $\lambda_{xx}^2 = 4^{1/d-1} D_{xx}/(2\pi n_0^{1/d})^2$. (iii): For a
crystalline system of independent electrons $\lambda_{xx}$ can be expressed
as a BZ integral. In three dimensions, for a cubic lattice and $m_b$
occupied bands, we can prove that: \begin{equation} \lambda_{xx}^2 = - \,
\frac{4^{-2/3}}{(2\pi n_0^{1/3})^2} \, \lim_{N \rightarrow \infty} N^{-1/3}
\ln |z_N|^2 = \frac{1}{3 m_b} \Omega_{\rm I}, \end{equation} where
$\Omega_{\rm I}$ is the BZ integral of Ref.~\onlinecite{nota}.

In conclusion, we have shown how to unambiguously measure localization in
the ground state of a many-electron system. We have shown over a few
examples how to discriminate between an insulator and a metal without
actually looking at the excitation spectrum, simply scrutinizing electron
localization in the ground eigenstate. For the special case of an
insulating crystal of noninteracting electrons we measure nothing else than
the localization of the Wannier functions, whereas in the correlated and/or
disordered case our approach to localization is not related---to our best
knowledge---to any previously known theory~\cite{Kudinov}. Our work opens
the way to further advances and leaves several important issues open. We
mention just a few of them: effects (possibly qualitative) of long--range
interaction upon $\lambda$; use of different approximate many--body
wavefunctions (such as {\it e.g.} quantum Monte Carlo); role of $\lambda$
in the context of Anderson localization in disordered systems.

R. R. acknowledges useful discussions with N. Marzari, F. Mauri, and D. 
Vanderbilt, as well as with many participants in the 1998 workshop
``Physics of Insulators'' at the Aspen Center for Physics, where part of
this work was performed. Partly supported by INFM (PRA HTSC) and by ONR
(grant N00014-96-1-0689).

\end{document}